\RequirePackage[hyphens]{url}
\documentclass[12pt]{article}
\usepackage{pdflscape}
\usepackage{tikz,natbib,xfrac,amsmath,amsthm,booktabs,subfigure,anysize,hyperref,graphicx,float,tabularx,multirow}
\usepackage[margin=1in]{geometry}
\usepackage{xcolor}
\usepackage{bm}
\definecolor{dark-red}{rgb}{0.4,0.15,0.15}
\definecolor{dark-blue}{rgb}{0.15,0.15,0.4}
\definecolor{medium-blue}{rgb}{0,0,0.5}
\definecolor{ChadBlue}{rgb}{.1,.1,.5}
\definecolor{ChadDarkBlue}{rgb}{.1,0,.2}
\definecolor{ChadBlue}{rgb}{.1,.1,.5}
\definecolor{ChadRoyal}{rgb}{.2,.2,.8}
\definecolor{ChadGreen}{rgb}{0,.4,0}    
\definecolor{ChadRed}{rgb}{.5,0,.5}  
\hypersetup{
    colorlinks, linkcolor={ChadRed},
    citecolor={ChadBlue}, urlcolor={medium-blue}
}
\newcommand\posscite[1]{\citeauthor{#1}'s (\citeyear{#1})}
\usepackage[font=small,labelfont=bf,justification=justified,singlelinecheck=false]{caption}
\usepackage{amssymb}
\usepackage{amsfonts}
\usepackage{amsmath}
\usepackage{setspace}  
\usepackage[all]{hypcap}        

\graphicspath{{figures/}}
\usepackage[english]{babel}
\usepackage{longtable}
\usepackage[singlelinecheck=off]{caption}
\usepackage[singlelinecheck=false]{caption}
\usepackage{array}
\usepackage{float}

\usepackage [autostyle, english = american]{csquotes}
\usepackage{chngcntr}
\usepackage{rotating, graphicx}
\usepackage{epstopdf}
\usepackage{threeparttable}
\usepackage{sectsty}
\sectionfont{\large}

\usepackage{longtable}
\usepackage{booktabs}
\usepackage{array}
\newcolumntype{L}{>{\raggedright\arraybackslash}p{1cm}}
\newcolumntype{K}{>{\raggedleft\arraybackslash}p{1cm}}
\newcolumntype{R}{>{\raggedright\arraybackslash}p{3cm}}
\newcolumntype{Q}{>{\raggedright\arraybackslash}p{10cm}}
\usepackage{tabu}
\usepackage{dcolumn}
\usepackage{ifpdf}

\title{Who presents and where? An analysis of research seminars in US economics departments\thanks{I thank Benedikt Heid, Nagore Iriberri, Asier Mariscal, Francisco Requena, and participants at the 2019 Simposio de An\'{a}lisis Econ\'{o}mico in Alicante for very valuable comments and suggestions. I am grateful to Christian Zimmermann for granting me access to the RePEc database. I gratefully acknowledge financial support from the Spanish Ministry of Science, Innovation and Universities (RTI2018-100899-B-I00, co-financed with FEDER) and the Basque Government Department of Education, Language policy and Culture (IT885-16).}}

\author{\large {Asier Minondo}\thanks{Deusto Business School, University of Deusto, Camino de Mundaiz 50, 20012 Donostia - San Sebasti\'{a}n (Spain). Research affiliate of Instituto Complutense de Estudios Internacionales. Email: \href{mailto:aminondo@deusto.es}{aminondo@deusto.es}}}

\date{ \today \\  }

\usepackage[normalem]{ulem}

\begin{document}

\maketitle

\begin{abstract}
Using a large dataset of research seminars held at US economics departments in 2018, I explore the factors that determine who is invited to present at a research seminar and whether the invitation is accepted. I find that high-quality scholars have a higher probability of being invited than low-quality scholars, and researchers are more likely to accept an invitation if it is issued by a top economics department. The probability of being invited increases with the size of the host department. Young and low-quality scholars have a higher probability of accepting an invitation. The distance between the host department and invited scholar reduces the probability of being invited and accepting the invitation. Female scholars do not have a lower probability of being invited to give a research seminar than men.
\end{abstract}

\begin{flushleft}
\end{flushleft}

\newpage 
\onehalfspacing

\section{Introduction}
To develop new ideas, scholars need to be aware of the state-of-the-art in the field and discuss their new research projects with peers. Research seminars contribute to the achievement of both objectives. Thus, university departments devote a considerable amount of resources to organize research seminars, and scholars invest a substantial share of time in preparing presentations and traveling to other universities. However, despite its being a core activity of academic life, little is known about the variables determining who is invited to deliver a research seminar and whether a scholar will accept the invitation to present. 

I build a large dataset of research seminars held at US economics departments in 2018. In my sample, I only observe whether a scholar gives a seminar at an economics department. I do not observe whether a department invites a scholar to deliver a seminar or whether a scholar accepts the invitation, but the product of these individual decisions. Despite this limitation, I can estimate the variables governing the decision to invite and accept using a bivariate probit model with partial observability. I find that high-quality scholars are more likely to be invited to deliver a research seminar and that scholars are more likely to accept an invitation to give a seminar if it is issued by a top department. These results suggest a positive assortative matching in seminars between the quality of the inviting department and the quality of the invited scholar. This matching may reinforce the quality advantage of ex-ante high-quality scholars and departments. Women do not have a lower probability to be invited to deliver a research seminar than men. Scholars are more likely to be invited by a large department. A long distance between the inviting department and affiliation of the invited scholar reduces the probability of issuing and accepting an invitation to deliver a research seminar. Low-quality scholars have a higher probability to accept an invitation than high-quality scholars. Young scholars are more willing to accept an invitation than senior scholars. 

To the best of my knowledge, the paper is the first to provide evidence on the variables that determine (i) the scholars invited to deliver a research seminar, and (ii) where a scholar is more likely to present a research seminar. Previous papers concluded that workshops and conferences facilitate the transmission of knowledge \citep{iaria2018frontier,deleon2018conferences,head2019geography}, promote collaboration among scholars \citep{campos2018conferences,chai2019conferences}, increase the probability of publishing in high-quality journals \citep{gorodnichenko2019conferences}, and the quality of research \citep{minondo2020comments}. I add to the literature by showing that high-quality departments are more likely to be aware of new knowledge generated in the field, because scholars producing the most promising ideas are more willing to present their new projects to such departments. Previous studies on urban economics, economic growth, and economics of innovation have shown that social interactions enable people to be exposed to new ideas, thus raising knowledge and fostering the development of new ideas \citep{lucas2009ideas,glaeser2011triumph,delaroca_puga2017learning,akcigit2018dancing,andrews2019bars}. Research seminars provide scholars with exposure to new ideas and facilitate interaction among peers. I contribute to the literature by analyzing the variables that raise the probability that a research seminar takes place and thus, foster awareness about new ideas and interaction among scholars. In general, this paper is related to the literature that explores the productivity and quality determinants of economics scholars and departments \citep{kim2009elite,bosquet2017sorting,hamermesh2018citations}. I contribute to this literature by showing that research seminars may further enhance the productivity of high-quality scholars affiliated with high-quality departments, because such scholars will have more opportunities to be aware of new ideas and improve their research projects on the basis of the comments and suggestions from high-quality peers \citep{minondo2020comments}. Finally, I join several studies in analyzing gender discrimination in economics \citep{ginther2004women,bayer2016diversity,lundberg2019women,hengel2019gender,card2020refereesgenderneutral}. I add to the literature showing that female scholars do not have a lower probability to be invited to deliver a research seminar than men.

The remainder of the paper is organized as follows. Section~\ref{sec:model} posits a simple analytical framework to understand the factors that motivate departments to organize research seminars and scholars to present at a research seminar. Section~\ref{sec:data} describes the dataset and presents some summary statistics. Section~\ref{sec:regressions} discusses the results of the regression analyses, and Section~\ref{sec:conclusions} concludes.

\section{Simple analytical framework on research seminars}
\label{sec:model}

To guide the empirical analysis I posit a very simple analytical framework to explain the factors that determine the scholars invited to deliver a research seminar and the reasons that lead a scholar to accept such an invitation. To produce high-quality research, university departments should be aware of new ideas, methodologies and databases in their field. A strategy for remaining at the frontier of knowledge is inviting the scholars who are generating new ideas and methodologies and using such novel databases. Research seminars are especially helpful for increasing awareness of cutting-edge research, because the presenters may have yet to publish a working paper of the new research project, and the new knowledge may still be tacit. Even when a working paper exists, oral presentations and discussions between the presenter and attendees may clarify certain aspects of the paper \citep{chai2019conferences}.

A department holds a limit on the number of research seminars it can host during an academic course; thus, it will aim to maximize the research quality of the presenters given the budget. I assume that departments only observe the quality of the scholar and give freedom to decide the paper that will be presented. I define a latent variable $I^*_{ds}$, which measures the willingness of department $d$ to invite scholar $s$ to deliver a research seminar. Analytically:

\begin{equation}
\label{eq:invites}
\begin{split}
I^*_{ds}=\beta_{1} Q_{s}+ \beta_{2} Cost_{ds}+  \beta_{3} Size_{d}+ \beta_{4} Female_{s}+ \epsilon_{ds}
\end{split}
\end{equation}

In addition to researcher's quality, $Q_{s}$, other factors may also determine a department's willingness to invite a scholar. For example, departments may be less willing to invite a scholar if travel expenses are high, $Cost_{ds}$. Large departments, $Size_{d}$, are more likely to invite scholars because they have more budget to invite speakers than small departments. This later variable is positively correlated with the quality of the department. Therefore, it also capture the fact that high-quality departments are more willing to invite because their invitations are more likely to be accepted. In contrast, low-quality department may decide not to issue invitations because they anticipate that researchers will decline the invitation.\footnote{As explained below when discussing the partial bivariate model, using the size of the department rather than the quality of the department allow us to introduce variability in the variables determining the invite and accept decisions.}

A recent survey of the American Economic Association indicated that 32\% of women, as opposed to 13\% of men, felt discriminated or unfairly treated in terms of being invited to participate in research conferences, associations and networks \citep{aea2019climate}. \cite{chari2017nber} found that the rate of paper acceptance at the NBER Summer Institute for women is statistically indistinguishable from that for men. However, \cite{hospido2019conferences} found that female-authored papers are less likely to be accepted at three major academic conferences in economics.\footnote{Regarding research seminars, The Econ Seminar Diversity project is gathering a database on who spokes at economics departments seminar series, and it provides a tool to visualize the percentage of women and scholars belonging to minorities that are invited to give a seminar. Available at \url{https://econseminardiversity.shinyapps.io/EconSeminarDiversity/}} To capture potential gender discrimination when inviting a speaker, I introduce a dummy variable, $Female_{s}$, which takes a value of one if the speaker is female and zero otherwise. $\epsilon_{ds}$ is the disturbance term. A department will deliver an invitation to present a seminar if the willingness to invite exceeds a given threshold $\lambda$. Thus, the probability that department $d$ invites scholar $s$ is $\mathbb{P}(I^*_{ds}>\lambda)$.

A scholar wants to present her research to high-quality audiences, where she is more likely to receive suggestions and comments that may enable her to improve the quality of a new project. Therefore, the probability that an author accepts an invitation to deliver a research seminar will be high with a high-quality department issuing the invitation. Low-quality scholars are more willing to accept an invitation than high-quality scholars, because their opportunities to present at a research seminar are lower. Scholars may be less willing to accept an invitation if it involves a long trip. Finally, young scholars may be more willing to accept an invitation because they want to present themselves to the research community \citep{chai2019conferences}. 

I define a latent variable $A^*_{sd}$, which measures the willingness of scholar $s$ to accept an invitation to present a paper at department $d$. Analytically, $A^*_{sd}$ can be expressed as follows:

\begin{equation}
\label{eq:accepts_invitation}
A^*_{sd}=\beta_{1} Q_{d}+\beta_{2} Q_{s}+\beta_{3} Trip_{sd}+ \beta_{4} Age_{s} + \epsilon_{sd}
\end{equation}

where $Q_{d}$ denotes the quality of the inviting institution, $Trip_{sd}$ the duration of the trip, $Age_{s}$ the career age of the scholar, and $\epsilon_{sd}$ the disturbance term. Scholar $s$ will accept the invitation to deliver a research seminar at department $d$ if the willingness to accept overcomes a given threshold $\kappa$. Thus, the probability of accepting an invitation is $\mathbb{P}(A^*_{sd}>\kappa)$.

A research seminar will take place if department $d$ delivers an invitation to scholar $s$, and scholar $s$ accepts the invitation.\footnote{The process can flow in the opposite direction: a scholar may offer to deliver a seminar and the department may accept the offer.} Thus, the probability of holding a research seminar by scholar $s$ at department $d$, $\mathbb{P}(S_{sd})$, can be expressed as follows:                                                       

\begin{equation}
\label{eq:probability_seminar}
\mathbb{P}(S_{sd})=\mathbb{P}(I^*_{ds}>\lambda, A^*_{sd}>\kappa)
\end{equation}

The model has two binary outcomes, namely, $I_{ds}$, which takes a value of one if  $I^*_{ds}>\lambda$, and zero otherwise, and, $A_{sd}$, which takes a value of one if $A^*_{sd}>\kappa$, and zero otherwise. However, my data only allow me to observe the product of these outcomes. I note that an invitation was issued if the author accepts it; and I only observe the absence of a match. In the latter case, I cannot determine whether the seminar did not occur because the department did not issue the invitation or the scholar did not accept the invitation. Following \cite{poirer1980bivariate}, the partial observability problem can be represented by a single binary random variable:

\begin{equation}
\label{eq:Z_definition}
Z_{ds}=I_{ds}A_{sd}
\end{equation}

The distribution of $Z_{ds}$ is

\begin{equation}
\label{eq:Z_distribution}
\begin{split}
p_{ds}=\mathbb{P}(Z_{ds}=1)=\mathbb{P}(I_{ds}=1 \text{ and } A_{sd}=1)=F(x_{I}\beta_{I},x_{A}\beta_{A};\rho), \\
1-p_{ds}=\mathbb{P}(Z_{ds}=0)=\mathbb{P}(I_{ds}=0 \text{ or } A_{sd}=0)=1-F(x_{I}\beta_{I},x_{A}\beta_{A};\rho)
\end{split}
\end{equation}

where $F$ denotes the bivariate standard normal distribution and $\rho$ the correlation of the error terms (i.e., $\epsilon_{ds}$ and $\epsilon_{sd}$). $x_{I}\beta_{I}$ and $x_{A}\beta_{A}$ are the variables and parameters included in Equations~\eqref{eq:invites} and~\eqref{eq:accepts_invitation}, respectively. 

The log-likelihood function of the sample is expressed as follows:

\begin{equation}
\label{eq:loglikelihood}
\begin{split}
L(\beta_{I},\beta_{A},\rho)=\sum_{d=1}^{n}\sum_{s=1}^{n}Z_{ds}\ln[F(x_{I}\beta_{I},x_{A}\beta_{A};\rho)]   \\
+(1-Z_{ds})\ln[1-F(x_{I}\beta_{I},x_{A}\beta_{A};\rho)]
\end{split}
\end{equation}

\cite{poirer1980bivariate} showed that $\beta_{I}$ and $\beta_{A}$ can be estimated if, at least, one variable included in one of the variable vectors, $x_{I}$ or $x_{A}$, is excluded from the other variable vector. In my model, $Size_{d}$ and $Female_{s}$ are included in $x_{I}$, but excluded from $x_{A}$. And, $Age_{s}$, included in $x_{A}$, is excluded from $x_{I}$  Hence, estimating all parameters included in $\beta_{I}$ and $\beta_{A}$ with a bivariate probit model with partial observability is possible.

The partial bivariate model assumes a correlation between the errors terms (i.e., $\epsilon_{ds}$, $\epsilon_{sd}$) in Equations~\eqref{eq:invites} and~\eqref{eq:accepts_invitation} This assumption is reasonable in the said context. For example, as explained above, a department may not issue an invitation if it is very unlikely that a scholar will accept the invitation. Therefore, willingness to accept an invitation ($\kappa$) is a factor that can determine the willingness to issue an invitation ($\lambda$), which leads to a correlation between errors terms. 

I argue that the regression equation incorporates the variables critical for the decisions that I am modeling. Obviously, other variables may also affect the probability of holding a seminar. For example, if a scholar is ill, she will be unable to accept the invitation to deliver a seminar. Conversely, if a department is located in an area that suffered an earthquake, it may be unable to host seminars until the faculty buildings are repaired. However, these factors are orthogonal to the variables included in the model, and their omission should not affect the estimates. 

Identification in bivariate probit models with partial observability is weaker than that in bivariate models with full information about individual decisions \citep{meng1985bivariateprobitpartial}. This is because the model has to estimate the parameters of two decisions from events that are incompletely observed. To test the robustness of results, I estimate a univariate probit model which combines the variables that affect the probability of inviting and accepting an invitation. The regression equation is expressed as follows:

\begin{equation}
S^*_{sd}=\beta_{1} Q_{s}+ \beta_{2} Q_{d}+ \beta_{3}  Cost_{ds}+ \beta_{4} Female_{s}+\beta_{5} Size_{d}+ \beta_{6} Age_{s}+ \epsilon_{ds}
\end{equation} 

Scholar $s$ will hold a seminar at department $d$ ($S_{sd}=1$) if the latent variable $S^*_{sd}$ is higher than a given threshold $\mu$. If the latent variable does not overcome such a threshold, then seminar will not take place ($S_{sd}=0$).

\section{Data}
\label{sec:data}
I randomly selected 143 economics departments out of the 240 economics programs included in Table~1 of \cite{mcpherson2012ranking}. The randomly selected departments for the current study are listed in Table~\ref{tab:seminar_list} in the Appendix.\footnote{I sampled up to 157 departments, but 11 of them could not be included in the sample due to limitations in retrieving the required information, and 3 of them because the number of professors in the department was less than five.} From the departments' web pages, I extracted information about seminars held in 2018, such as the invited scholar, affiliation of the invited scholar, and, if available, the title of the paper that was presented. I excluded from the sample the seminars given by scholars that belonged to the department.\footnote{I also excluded the recruitment seminars, or the seminars delivered by students that were doing their doctoral studies in the department.} I include a department in the sample even if it did not hold any research seminar in 2018. The sample of scholars is composed of professors affiliated with the 240 US economics departments included in Table~1 of \cite{mcpherson2012ranking}.\footnote{I do not include emeritus professors or joint appointments.} Following \cite{bosquet2019genderandpromotions}, I exclude from the sample the economics departments with less that five professors. The estimation sample is generated by crossing the randomly selected departments with the sample of scholars, and excluding the combinations where the potential speaker is a member of the department.

I use the index elaborated by \cite{mcpherson2012ranking} to proxy for the quality of US economics departments. This index is based on the number of pages published by a department's scholars in the top 50 economics journals during 2002-2009. I use two measures to proxy for the quality of the invited scholar, namely, (i) quality of the economics department to which the scholar is affiliated; and (ii) number of citations to the scholar's research outputs according to her profile in Google Scholar, adjusted by the career age of the scholar. I use the same variable to capture the cost of inviting a scholar, $Cost_{ds}$, and length of the trip, $Trip_{sd}$, that is, the distance between the inviting department and affiliation of the invited scholar.\footnote{I calculate bilateral distances using the latitude and longitude of the inviting department and affiliation of the invited scholar.} To calculate the career age of the scholar, I identify her earliest publication in Google Scholar. I calculate career age as 2018 minus the year in which the earliest work was published, plus one. I measure the size of an economics department by the number of professors affiliated with it.

\begin{table}[t]
	\begin{center}
		\caption{Building of sample and summary statistics}
		\label{tab:summary_statistics}
\begin{tabular}{l r}
\toprule
Number of US economics departments in the random sample&143\\
\hspace{0.5cm}  which held at least one research seminar in 2018&86\\
Number of scholars affiliated with a US economics department&4,844\\ 
\hspace{0.5cm} Female&1,119\\
\hspace{0.5cm} With a Google Scholar ID&2,940\\
Number of observations in the estimation sample&689,766\\
Number of seminars&1,553\\
Number of scholars presenting at a seminar&936\\
 \hspace{0.5cm} Female&186\\
\midrule
\end{tabular}
\small
\begin{tabular}{l r r r r r}
&Median&Mean&Std. dev.&Min&Max\\
\# Seminars per department& 3.0&10.9&    16.8&          0   &     78\\
\# Seminars per scholar&0.0&0.3&0.8&0&7\\
\bottomrule
\end{tabular}
		\caption*{\hspace{1.8cm}\begin{footnotesize} Note: Author's calculations.
		\end{footnotesize}}
	\end{center}
\end{table}

Table~\ref{tab:summary_statistics} provides the summary statistics of the sample. The number of US economics departments included in our sample is 143, out of which 86 held at least one seminar in 2018. The number of scholars affiliated with a US economics department was 4,844, out of which 1,119 were female (23\%). The number of scholars affiliated with a US economics department that had a Google Scholar ID was 2,940 (61\%). The crossing of the randomly sampled departments (143) and scholars affiliated with US economics departments (4,844), minus the scholars affiliated to the randomly selected departments (2,926) generates an estimation sample of 689,766 observations. I retrieved data from 1,553 seminars held by economics departments included in the sample.\footnote{I retrieved information on 3,943 seminars. Many seminars were not included in the estimation sample because they were given by scholars affiliated with a non-US university, with the department, or with business schools, law schools, other non-economics departments in US universities, or other US institutions (e.g. federal reserve banks).} The number of scholars presenting at least one research seminar was 936. A total of 186 of the latter speakers were female (20\%). 

The bottom panel of Table~\ref{tab:summary_statistics} indicates that the median department held 3 seminars in 2018. However, the distribution of seminars per department was highly skewed, because the average number of seminars was 11 and the standard deviation 17. A total of 57 departments did not hold any seminar in 2018, whereas one department held up to 78 seminars. The median number of seminars per scholar was zero. This distribution was also skewed as indicated by the average number of seminars per scholar of 0.3 and a standard deviation of 0.8. In total, 3,908 scholars did not present at a research seminar, whereas some scholars presented at seven economics departments in 2018.

\begin{figure}[t!]
	\begin{center}
		\caption{Correlation between quality and number of seminars, 2018}
		\label{fig:correlation}
		\begin{tabular}{c}
			A. Economics department\\
		\includegraphics[height=3.5in]{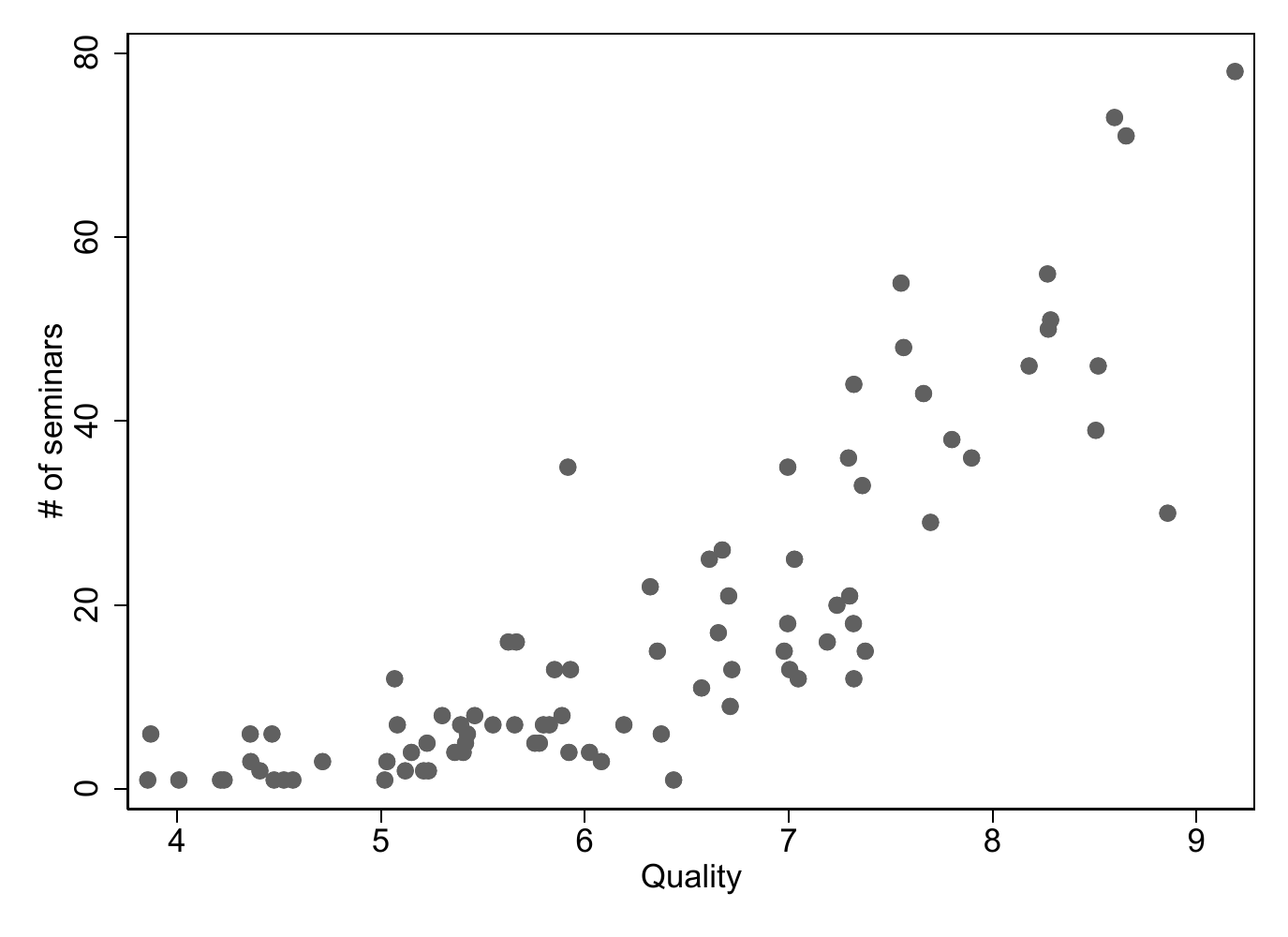}\\
			B. Scholar\\
			\includegraphics[height=3.5in]{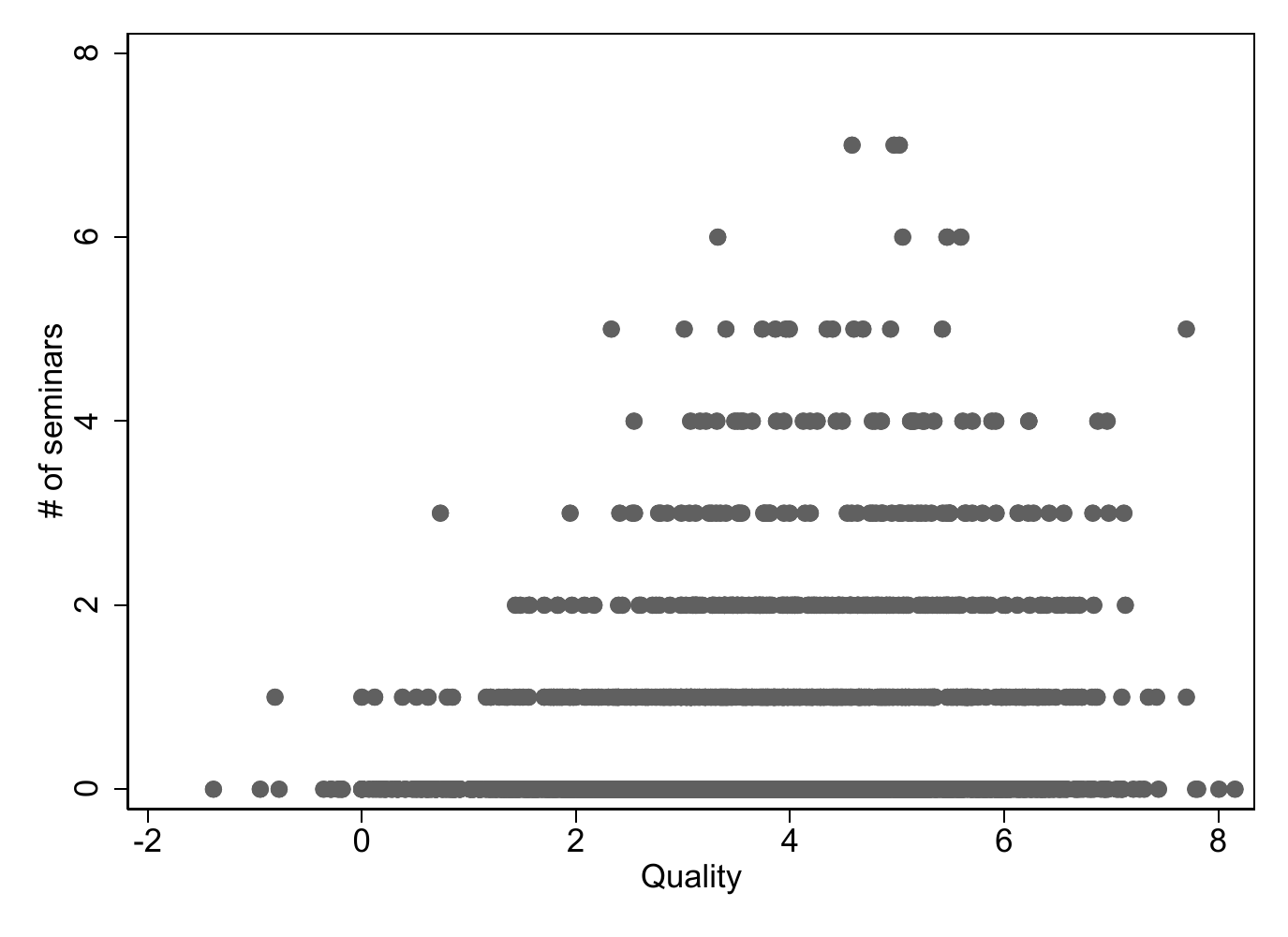}\\\end{tabular}
			\end{center}
\footnotesize{Note: The quality of economics departments is measured by \posscite{mcpherson2012ranking} index (in logs). The quality of scholars is measured by the (log) number of citations to works recorded by Google Scholar divided the career age.}
\end{figure}

Panel A in Figure~\ref{fig:correlation} shows a positive correlation between the quality of the economics department, measured by \posscite{mcpherson2012ranking} ranking, and number of research seminars held by that department. The number of research seminars increases exponentially with the quality of the economics department. Panel B in Figure~\ref{fig:correlation} plots the correlation between the quality of scholars, measured by the number of citations in Google Scholar adjusted by the career age, and number of seminars delivered in 2018. The vertical axis measures the number of seminars, which ranges from 0 to 7 (Table~\ref{tab:summary_statistics}). Each dot in the graph corresponds to a scholar. For each number of seminars, the quality-range of scholars is very wide. However, the average quality of scholars increases (i.e., the line of dots moves to the right) as the number of seminars delivered by a scholar increases. These figures suggest that research seminars are more likely to occur if the quality of the host department and invited scholar is high. The regression analyses carried out in the next section analyze whether this visual appreciation is correct.

\section{Regression results}
\label{sec:regressions}

This section presents the results of econometric analyses. To maximize the number of observations, I first proxy for the quality of the invited scholar by the quality of the department to which she is affiliated. The shortcoming of this estimation is that I cannot include career age as independent variable because it can only be computed for authors with a Google Scholar profile.

Column (1) of Table~\ref{tab:mcpherson} presents the baseline estimation. Standard errors are clustered at the scholar level. At the bottom of the table, I report the chi-square statistics for the likelihood ratio test that the correlation between the error terms is zero. The null hypothesis of the absence of correlation is strongly rejected. The probability that a scholar is invited to deliver a research seminar increases with the quality of the department to which she is affiliated. This result is in line with the prediction that departments seek to invite high-quality scholars to deliver a seminar. In turn, scholars are more likely to accept an invitation if it is issued by a high-quality department. This result is also in line with the prediction that scholars aim to present their papers to high-quality audiences. Gender (i.e., being a female) does not reduce the probability of being invited to give a seminar. Larger departments have a higher probability to invite than smaller departments. Distance has a negative effect on the probability to invite and accept. Finally, scholars affiliated with low-quality universities are more willing to accept an invitation to give a seminar.

\begin{table}[t!]
	\begin{center}
		\caption{Probability to invite and accept}
		\label{tab:mcpherson}
{
\def\sym#1{\ifmmode^{#1}\else\(^{#1}\)\fi}
\begin{tabular}{l*{4}{c}}
\toprule
&\multicolumn{4}{c}{Scholar quality proxied by}\\
     &\multicolumn{2}{c}{Affiliation}&\multicolumn{2}{c}{Citations}\\
     \cmidrule(lr){2-3}  \cmidrule(lr){4-5}               
                    &\multicolumn{1}{c}{(1)}&\multicolumn{1}{c}{(2)}&\multicolumn{1}{c}{(3)}&\multicolumn{1}{c}{(4)}\\
                    &\multicolumn{1}{c}{All}&\multicolumn{1}{p{3.25cm}}{\centering Active hosts \\ \&presenters}&\multicolumn{1}{c}{All}&\multicolumn{1}{p{3.25cm}}{\centering Active hosts \\ \&presenters} \\
\midrule
\textit{Probability to invite}&                   &                   &                   &                   \\
Scholar quality     &       0.443\sym{a}&       0.289\sym{a}&       0.150\sym{a}&       0.109\sym{a}\\
                    &     (0.018)       &     (0.017)       &     (0.025)       &     (0.024)       \\
[1em]
Department size     &       0.004\sym{a}&       0.003\sym{a}&       0.002       &       0.003\sym{b}\\
                    &     (0.001)       &     (0.001)       &     (0.002)       &     (0.001)       \\
[1em]
Female              &       0.016       &       0.018       &      -0.078       &       0.007       \\
                    &     (0.044)       &     (0.025)       &     (0.072)       &     (0.014)       \\
[1em]
Distance            &      -0.098\sym{a}&      -0.117\sym{a}&      -0.070\sym{a}&      -0.090\sym{a}\\
                    &     (0.012)       &     (0.015)       &     (0.021)       &     (0.027)       \\
\midrule
\textit{Probability to accept}&                   &                   &                   &                   \\
Department's quality&       0.646\sym{a}&       0.796\sym{a}&       0.527\sym{a}&       0.149\sym{c}\\
                    &     (0.072)       &     (0.109)       &     (0.080)       &     (0.090)       \\
[1em]
Scholar quality     &      -0.167\sym{c}&      -0.871\sym{a}&       0.052       &      -0.136\sym{a}\\
                    &     (0.086)       &     (0.069)       &     (0.052)       &     (0.037)       \\
[1em]
Distance            &      -0.143\sym{a}&      -0.043       &      -0.158\sym{a}&       0.033       \\
                    &     (0.017)       &     (0.047)       &     (0.027)       &     (0.046)       \\
[1em]
Career age         &                   &                   &      -0.025\sym{a}&      -0.002       \\
                    &                   &                   &     (0.004)       &     (0.001)       \\
\midrule
Log Pseudolikelihood&   -8762.424       &   -6961.114       &   -6914.602       &   -5460.510       \\
Observations        &      689766       &       94337       &      386447       &       66821       \\
Test $\rho$=0       &      21.767       &       5.067       &       0.890       &       9.979       \\
\bottomrule
\end{tabular}
}

		\caption*{\begin{footnotesize}Note: Constants are not reported. Standard errors clustered at the scholar level are in parentheses. a, b, c: statistically significant at 1\%, 5\%, and 10\%, respectively.\end{footnotesize}}
	\end{center}
\end{table}

I use the coefficients in column~(1) to quantify the changes in the probability of being invited and accepting the invitation as I vary the value of the variables included in the model. For example, the probability that a (male) scholar affiliated with Yale, a department that occupies the 10th position in the quality ranking, is invited to deliver a seminar at Stanford is 242 times larger than if the scholar was affiliated with Richmond, a university at a similar distance from Stanford, but which occupies the 194th position in the quality ranking. Likewise, the probability that a scholar affiliated with Stanford accepts an invitation from Yale is 3,340 times larger than the probability of accepting an invitation from Richmond. A scholar affiliated with Berkeley has a 2.5 times higher probability to be invited by Stanford than a scholar from MIT, a department that has a similar quality-ranking as Berkeley, but is much farther from Stanford. Likewise, a scholar from Berkeley has a 2.7 times higher probability of accepting an invitation from Stanford than a scholar from MIT. 

To test the robustness of results, I remove the departments without seminars and scholars that did not present at any seminar from the sample. Column (2) of Table~\ref{tab:mcpherson} presents the results. Except for distance in the accept decision, the estimates are qualitatively similar to those obtained in the baseline analysis. As expected, a drop occurred in the value of scholar's quality and department's quality coefficients due to the sample selection. However, these coefficients remain positive and precisely estimated.  

In the second set of estimations, I measure the quality of the invited scholar by the number of citations to her work. I divide this figure by the career age of the scholar to control for the fact that scholars with longer careers receive more citations than junior scholars. This measure is deemed a better proxy for the quality of a scholar than her affiliation because of heterogeneity in the quality of scholars, as measured by citations, within economics departments \citep{hamermesh2018citations}. The information retrieved from the Google Scholar profiles also enables me to include career age as additional explanatory variable to the decision of accept. The ``cost" of using these new data is a sizable reduction in the number of observations.

Column~(3) of Table~\ref{tab:mcpherson} reports the baseline estimates. The chi-square coefficient reported at the bottom of the table is low. Hence, we cannot reject the null hypothesis of absence of correlation between the invitation and accept decisions' error terms. A high-quality scholar has a larger probability of being invited to present a research seminar than a low-quality scholar. In turn, scholars are more likely to accept an invitation if it is issued by a high-quality department. These results confirm the positive assortative matching in research seminars between high-quality scholars and high-quality departments.

Large departments are more willing to invite scholars, but the coefficient is statistically not significant. Distance has a negative effect on the probability to invite and accept, although the impact is stronger on the latter. Young scholars are more willing to accept an invitation than senior scholars. The female coefficient is negative, but statistically not significant. There is no indication that high-quality scholars are less willing to accept an invitation to give a seminar.

I use the coefficients in column~(1) of Table~\ref{tab:mcpherson} to explore how variations in the value of the independent variables alter the probability of inviting and accepting an invitation. For example, a male scholar with a median career age (18 years), affiliated with an east-coast university (for example, Yale), and at the top 80\% of the age-adjusted citation distribution has a 2.3 times higher probability to be invited by Stanford than a male scholar, with the same career age, affiliated with the same east-coast university, but located at the top 20\% of the citation distribution. Likewise, a male scholar affiliated with a west-coast university (for example, Stanford), a median career age, and at the top 80\% of the five-year citation distribution is 113 times more likely to accept an invitation to give a seminar at Yale than at Richmond. A male scholar at the top 80\% who is affiliated with Berkeley has a 2.5 times higher probability of being invited to deliver a seminar at Stanford than a similarly ranked male scholar from MIT. Likewise, a male scholar from Berkeley has a 1.9 times higher probability to accept an invitation from Stanford than a male scholar from MIT. Finally, the probability that a top 80\% scholar from Stanford delivers a seminar in his department is 1.8 times higher than if he were affiliated with Berkeley.

I also test the robustness of results removing the departments that did not hold any seminar and scholars that did not deliver any seminar in 2018 from the sample (column~(4)). As expected, the quality coefficients for the presenter and host department have lower values than those in the baseline estimation (column~(3)), but remain positive and statistically significant. Distance reduces the probability to invite, but does not affect the probability to accept. Low-quality scholars have a larger probability to accept an invitation than high-quality scholars, and large departments have a larger probability to invite than small departments. Young scholars have a larger probability to accept an invitation than senior scholars, but the coefficient is statistically not significant.

\begin{table}[t]
	\begin{center}
		\caption{Univariate probit. Probability of holding a seminar}
		\label{tab:univariate}
		{
\def\sym#1{\ifmmode^{#1}\else\(^{#1}\)\fi}
\begin{tabular}{l*{2}{c}}
\toprule
                    &\multicolumn{1}{c}{(1)}&\multicolumn{1}{c}{(2)}\\
                    &\multicolumn{1}{c}{Affiliation}&\multicolumn{1}{c}{Citations}\\
\midrule
Scholar quality     &       0.253\sym{a}&       0.129\sym{a}\\
                    &     (0.010)       &     (0.015)       \\
[1em]
Department's quality&       0.287\sym{a}&       0.288\sym{a}\\
                    &     (0.011)       &     (0.013)       \\
[1em]
Department size     &       0.001       &       0.001       \\
                    &     (0.001)       &     (0.001)       \\
[1em]
Female              &       0.019       &      -0.050       \\
                    &     (0.030)       &     (0.037)       \\
[1em]
Distance            &      -0.105\sym{a}&      -0.120\sym{a}\\
                    &     (0.008)       &     (0.009)       \\
[1em]
Career age          &                   &      -0.015\sym{a}\\
                    &                   &     (0.001)       \\
[1em]
Own scholar         &                   &                   \\
                    &                   &                   \\
\midrule
Log Pseudolikelihood&   -9028.073       &   -6933.933       \\
Observations        &      689766       &      386447       \\
\bottomrule
\end{tabular}
}

		\caption*{\begin{footnotesize} \centering Note: Constants are not reported. Standard errors clustered at the scholar level are in parentheses. a, b, c: statistically significant at 1\%, 5\%, and 10\%, respectively. 			
		\end{footnotesize}}
	\end{center}
\end{table}

Identification in a partially observable biprobit model is weaker than in a model where both individual decisions are observable. To test the robustness of the results, I estimate a univariate probit model where the only decision is whether or not to hold a seminar (Equation~\eqref{eq:probability_seminar}). Table~\ref{tab:univariate} presents the results. In column~(1), I approximate the quality of a scholar by the quality of her affiliation. The quality of the scholar and economics department are positively correlated with the probability of holding a seminar. Distance between the invited scholar and inviting department reduces the probability of holding a seminar. The size of the department has a positive sign, but the coefficient is imprecisely estimated. The coefficient for female scholars is nearly zero, which indicates that gender does not have an effect on the probability of holding a seminar. These results are consistent with those obtained from the partially observable bivariate probit model (column (1) of Table~\ref{tab:mcpherson}).

Column~(2) presents the results of estimating Equation~\eqref{eq:probability_seminar} when the quality of the presenter is measured by the number of citations to her works adjusted by the age of the scholar. The probability of holding a seminar increases with the quality of the scholar and inviting department. A seminar has a higher probability of taking place when the scholar is young and if a short distance is observed between the invited scholar and inviting department. Large departments have a higher probability of holding a seminar than small departments, but the coefficient is imprecisely estimated. The coefficient for female scholars is negative, but statistically non-significant. These results are also consistent with those obtained from a partially observable bivariate probit model (column (3) in Table~\ref{tab:mcpherson}).

\section{Conclusions}
\label{sec:conclusions}
The paper explores the variables that determine who is invited to deliver a research seminar and where scholars want to present their new research projects. I find that the probability of being invited to present at a research seminar is positively correlated with the quality of the scholar, and scholars are more likely to accept an invitation if it is issued by a high-quality department. I also show that the geographical distance between departments and scholars reduces the probability of being invited and accepting the invitation. Large departments have a higher probability to invite than small departments, and young scholars are more willing to accept an invitation than senior scholars. Low-quality scholars have a larger probability to accept an invitation than high-quality scholars. Female scholars do not have a lower probability of being invited to deliver a research seminar than male scholars.

These results suggest that scholars affiliated with high-quality departments have more opportunities to listen to high-quality scholars and increase awareness of the state-of-the-art in the field, enabling to endow their new research with advanced ideas, methodologies, and databases. In turn, high-quality scholars have additional opportunities to improve their papers because they are more likely to receive comments and suggestions from other high-quality scholars. The positive assortative matching in seminars between high-quality departments and high-quality scholars provides top departments and scholars a tool to retain their leading positions. Low-quality departments and scholars can compensate for this disadvantage if they are located close to high-quality departments.

\appendix 
\label{app:all}

\setcounter{equation}{0}
\renewcommand\theequation{A.\arabic{equation}}
\setcounter{figure}{0}
\renewcommand\thefigure{A.\arabic{figure}}

\setcounter{table}{0}
\renewcommand\thetable{A.\arabic{table}}

\newpage

\scriptsize
\begin{longtable}{L R Q K}
	\caption{Economics departments included in the seminar sample} \label{tab:seminar_list} \\
	Ranking&Institution& Seminars included &\# of seminars\\	
	\midrule
	\endfirsthead
	
		\multicolumn{4}{l}%
		{\tablename\ \thetable\ -- \textit{Continued from previous page}} \\   
		\endfoot 
		\midrule
		Ranking&Institution&Seminars included&\# of seminars  \\
		\midrule
		\endhead
		\hline
		 \multicolumn{4}{l}{\textit{Continued on the next page}} \\
		\endfoot
		\bottomrule
		\endlastfoot
		\midrule
1&Harvard& The Monetary and Fiscal Policy Seminar; The Political Economy of Religion Seminar; The Public Economics and Fiscal Policy Seminar; The Law, Economics, and Organizations Workshop; Economics of Science and Engineering Workshop; The International Economics Workshop; The Program on Political Economy; Industrial Organization Workshop; Behavioral \& Experimental Economics Workshop, The Economic Development Workshop; Health Economics Workshop; The Theory Workshop; The Econometrics Workshop; The Economic History Workshop; The Labor Workshop; Seminar in Macroeconomic Policy; Seminar in Behavioral and Experimental Economics; Seminar in Public Economics and Fiscal Policy; Seminar in Economic Theory; Seminar in Econometrics; Seminar in Monetary and Fiscal Policy; Seminar in Industrial Organization; Seminar in Economic History; Seminar in Law, Economics, \& Organization; Seminar in Financial Economics; Seminar in Environmental Economics and Policy; Seminar in Labor Economics. Some seminars were organized jointly with MIT&211\\
2&Chicago&Applications Workshop; Econometrics Workshop; Money\&Banking Workshop; Workshop in Economic Theory; Workshop in Family Economics&73\\
4&MIT&Applied Microeconomics Seminar; Development Economics Seminar; Econometrics Workshop; Economics IAP; Finance Seminar; IO Workshop; International Seminar; Macro Seminar; Program on Political Economy; Public Finance/Labor Workshop; Seminar in Organizational Economics; Special Events; Theory Workshop. Some seminars were organized jointly with MIT&180\\
5&Stanford& Arrow Lectures; Department Seminar; Joint Applied Micro Seminar; Development; Econometrics; Experimental Behavioral Seminar; Economics Brown Bag Lunch Series; GSB Economic Theory; GSB Finance; GSB Organizational Behavior Seminars; GSB Political Economy; Industrial Organization; International Trade; Labor; Law and Economics; Macroeconomics; Public Economics and Environmental Economics; SIEPR Social Science and Technology&261\\		
7&Northwestern&Applied Micro Lunch; Development Economics Lunch Seminar; Development Lunch Seminar; Joint CET/CMS - EMS Theory Workshop; Macroeconomics Lunch Seminar; Seminar in Applied Microeconomics (Development, Labor and Public Economics); Seminar in Econometrics; Seminar in Economic History; Seminar in Industrial Organization; Seminar in Macroeconomics; Theory Bag Lunch&107\\
8&Penn&Econometrics Lunch; Econometrics Seminar; Empirical Micro Seminar; Industrial Organization Seminar; Macro Lunch; Micro Theory Lunch; Micro Theory Seminar; Money Macro Seminar&96\\
10&Yale&The Behavioral Sciences Workshop; Cowles Lunch Talks; Development Lunch; Development Workshop; Econometrics Seminar; Economic History Workshop; International Trade Lunch; Labor/Public Economics Workshop; Labor/Public Economics Prospectus Workshop; Industrial Organization Seminar; ISPS Event; Leitner Political Economy Seminar; Macro Lunch; Macroeconomics Workshop; Microeconomic Theory Workshop; Micro Theory Lunch; Partner Event; Simon Kuznets Lecture; Wasserman Workshop in Law and Finance;  YLS Center for the Study of Corporate Law &200\\
11&Michigan&Abraham and Thelma Zwerdling Lecture; Applied Microeconomics/IO Seminar; Causal Inference in Education Research Seminar; Econometrics; Economic Development Seminar; Economic History; Economic Theory; Health, History, Demography \& Development; International Economics; Interdisciplinary Seminar in Quantitative Methods; Labor Economics; Law and Economics Workshop; Macroeconomics; Public Finance; Social, Behavioral \& Experimental Economics; W.S. Woytinsky Lecture &136\\
12&Princeton&Behavioral Economics; CHW-RPDS; Griswold Center Event; Industrial Organization; Department Wide Seminars; Industrial Relations; International Trade; Macro/International Macro; Microeconomic Theory; Oskar Morgenstern Memorial Seminar; Political Economy Workshop; Simpson Lecture; Summer Seminar Series&131\\
13&UCLA&Albert Family Fund Seminar in Applied Microeconomics; Vongremp Workshop in Economic and Entrepreneurial History; Workshop in Econometrics; Workshop in Economic Theory; Laub Foundation Workshop in Industrial Organization; Ettinger Fund Workshop in Macroeconomics; Workshop in Trade, Economic Geography and Development &95\\
16&Maryland&Econometrics; IO/Theory; Labor/Public Finance/Development; Macroeconomics/International Finance; Trade/Institutions/Politics&84\\
18&UC San Diego&Applied Seminar Series; Econometrics Seminars Series; Int/Dev Seminar series; Macro Seminar Series; Metrics Seminar Series; Theory Seminar Series&81\\
19&Wisconsin-Madison&Robert E. Baldwin International Workshop; Joseph Krislov Labor Workshop; Juli Plant Grainger Econometrics Workshop; Juli Plant Grainger Industrial Organization Workshop; Juli Plant Grainger Macroeconomics Seminar; Juli Plant Grainger Public Workshop; Juli Plant Grainger Theory Workshop&64\\
21&Ohio State&Applied Microeconomics Seminar; Econometrics Seminar; Economic Theory/Experimental Seminar; Macroeconomics Seminar&69\\		
22&Minnesota&Agricultural and Applied Economics Seminar; Applied Micro; Department Seminar; Environmental and Resource Economics Seminar; Fed Bag Lunch; Finance Department Seminar; Jon Goldstein Memorial Lecture; Math Econ Seminar;  Micro-Macro Seminar; Minnesota Economics Seminar; Minnesota Lecture; MPC Seminar Series; Strategic Management and Entrepreneurship Seminar; Trade and Development Seminar&174\\
24&UC Davis&Behavioral; Development; Econometrics; Economic History; Energy; Environmental Economics; Industrial Organization; Macro/International Economics; Public Finance-Labor; Theory&128\\
26&Carnegie Mellon&Not available&\\
27&Dartmouth&Dartmouth IO Winter Conference; Economics Seminars; Household Finance Seminar; International Seminar&40\\
28&Rochester&Applied Workshop; International Workshop; Jones Lecture; Macro Workshop; McKenzie Lecture; Theory Workshop&47\\
30&Penn State&Applied Micro; Macroeconomics; Econometrics; Trade and Development; Micro Theory &85\\
31&Iowa State&Charles Sivesind Memorial Lecture; Department Seminars; George A. Fuller Memorial Lecture; I.W. Arthur Memorial Seminar; Pioneer Policy Lecture; William G. Murray Memorial Seminar &26\\		
32&North Carolina&Economics Seminars&60\\
34&Vanderbilt&Applied Economics; Departmental Macro; Departmental Micro; Econometrics; Economic History; Empirical Micro; Health; International; Political Economy&59\\
36&Boston College&Applied Microeconomics Seminar; Econometrics Seminar; Macroeconomics and Financial Economics Seminar; Macroeconomics Lunch; Microeconomics Seminar&85\\
38&UC Irvine&Econometrics Seminar; Labor-Public Seminar; Macroeconomics Seminar; Theory, History and Development Seminar; Transportation, Urban and IO Seminar&54\\
39&Purdue&Economics Seminar&33\\
45&Emory&Department-wide;Econometrics; Lunch\&Learn;Macroeconomics; Microeconomics&21\\
46&Arizona State&Economic Seminars&60\\
47&George Mason&ICES Experimental Economics Brown Bag Lecture; Micro-Economic Policy Seminar; Public Choice Seminar; Seminars; Washington Area Economic History Seminar; Workshop in Philosophy, Politics \& Economics&53\\
49&Pittsburgh&Seminars&96\\
50&Rutgers&Econometrics; Empirical Microeconomics; Macroeconomic Theory; Micro Theory/Experimental Seminar; Money, History and Finance&56\\
51&University of Washington&Econometrics; International Economics and Macroeconomics; Joint Seminar in Development Economics Series; Microeconomics&33\\
52*&Colorado&Could only retrieve data for Spring Series&\\
54&Iowa&No seminar series &\\
56&Georgia&Economics Seminar Series&24\\
57&North Carolina State&Macro Seminar Series; NCSU Econometrics Workshop; Microeconomics Workshop Series&18\\
58&Houston&Macroeconomics Series; Empirical Microeconomics Series&39\\
60&Rice&Brown Bag Seminars; Kalai Family Workshop in Applied Microeconomics; Kalai Family Workshop in Business and Economics; Kalai Family Workshop in Econometrics&48\\
61&UC Santa Cruz&Brown Bag Seminars; Macroeconomics \& International Finance Seminars; Microeconomics \& International Trade Seminars&50\\
62&Johns Hopkins&Seminars&54\\
64&Oregon&EC Seminar&17\\
68& Missouri&Brown Bag Seminar; Regular Seminar  &11\\
70&Brigham Young&$R^{2}$ Research; Visiting Scholar Seminar&20\\
72&Kentucky&Seminars and Workshops &27\\
73&Connecticut&Friday Econometrics Lunch; IO, Environmental, and Law Economics; Labor, Development, and Health Economics; Macroeconomics;&53\\
74&Texas-Dallas&No seminar series&\\
75&Claremont McKenna&RDS Seminar Series&11\\
76&Utah& No seminar series&\\
77&Wisconsin-Milwaukee&Seminars in the Center for Research on International Economics and the Department of Economics&19\\
81&Oregon State&Applied Economics Seminar Series&5\\
83*&Baruch College-CUNY&Could only retrieve data for the 2018 Fall Series&\\
87&Case Western&Economics Research Seminar&8\\
91&Oklahoma&Economics Research Seminar Series&23\\
93&Kansas&Seminars&8\\
94&UC Riverside&Applied Economics; Brown Bag; Econometrics; Economic Theory&71\\
96&Drexel&School of Economics Seminars&15\\
99&SUNY Albany&Seminars&25\\
100&Williams College&Economic Class of 1960 Scholars Seminar; Economics Department Seminar&24\\
102&Colorado-Denver&Seminars&12\\
103&American University&Research Seminar Series&27\\
106&Stony Brook&Departmental Research Series&28\\
109&South Carolina&No seminar series&\\
111&West Virginia&Economics Seminar Series&28\\
113&IUPUI&Economic Theory Workshop; Health Economics Seminar; Robert Sandy Economics Seminar&17\\
114&Auburn&Friday Seminar Series&27\\
117&Brandeis&Seminar Series&38\\
120*&Swarthmore&Not available&\\
121& Nevada-Las Vegas&No seminar series&\\
122&Middlebury&Economics Department Seminars&13\\
123&Mississippi&Seminar series&9\\
124&Nebraska&Economics Seminars&11\\
125&North Carolina - Charlotte&Economics Seminars Series&11\\
126*&Fordham&Not available&\\
127&Northeastern&Research Seminars&19\\
129&Cal State-Fullerton&Spring Seminar&4\\
130*&Graduate Center CUNY&&\\
132&San Diego State&No seminar series&\\
133&Florida Atlantic&No seminar series in 2018&\\
134&Texas Tech&Free Market Institute's Research Workshop&22\\
136&Texas Arlington&No seminar series in 2018&\\
137*&Vermont&Not available&\\
138&UNC-Greensboro&Economics Seminars&2\\
139&Wesleyan&No seminar series in 2018&\\
140&Bentley&No seminar series&\\
141&South Florida&Seminar Series&11\\
142&Cincinnati&Seminar Series&9\\
144& Miami-Ohio&Could not retrieve data for Sping 2018 seminar series&\\
145&Utah State& Seminars&16\\
146&Baylor&Seminars&6\\
147&Memphis&Seminar Series&13\\
148&Hawaii&Seminar Series&22\\
149&Temple& No seminar series&\\
150&Rhode Island&No seminar series&\\
152&Wake Forest&Seminars&12\\
154&North Texas&Department of Economics Seminar Series&2\\
156&Texas San Antonio&No seminar series&\\
158&Amherst& No seminar series&\\
161**&Nebraska-Omaha&No seminar series&\\
162&Illinois State&Economics Department Seminar Series;International Seminar Series;Applied Econometrics Workshop;Econometrics Workshop;Seminars sponsored by the Institute for Corruption Studies&15\\
163&Cal State-Sacramento&No seminar series&\\
165&Villanova&No seminar series&\\
166&Occidental College&No seminar series&\\
167&Union College&No seminar series&\\
168&Towson&Economics Department Seminar Series&4\\
171&Cal. Polytech State&Seminar&4\\
172&San Houston State&Seminar series&4\\
173&Middle Tennessee State&No seminar series&\\
174&New Hampshire&Not available&\\
175&Hamilton College&No seminar series&\\
176&Trinity University&No seminar series&\\
177&Loyola Marymount&Economics Seminar Series&7\\
178&Ohio&Economics Seminar&3\\
180&New Mexico&Graduate Seminars&1\\
181&North Dakota&Economic Seminar Series&10\\
184&Lafayette College&No seminar series&\\
185&Texas Christian&No seminar series&\\
186*&St. Louis &&\\ 
187&Lehigh&Department of Economics Seminar Series&6\\
188**&Colby College&Seminars&7\\
189&Northern Illinois&Economics Seminar Series&13\\
190&Cal. State-Northridge&No seminar series&\\
191&North Dakota State&No seminar series&\\
193&Old Dominion&No seminar series&\\
194&Richmond&No seminar series&\\
195&Dayton& No seminar series&\\
196&Kenyon College&No seminar series&\\
197&Akron&No seminar series in 2018 &\\
198&Washington and Lee&W\&L/VMI Seminars&7\\
199&Air Force Academy&No seminar series&\\
200&Portland State&Economics Seminar Series&10\\
202&Gettysburg College&No seminar series&\\
203*&Queens College&Not available &\\
204*&Missouri-St. Louis&Not available&\\
205&Saint Cloud State& No seminar series&\\
206*&Smith College&Not available&\\
207&Barnard College&No seminar series&\\
208&Chapman&No seminar series&\\
209*&Clark University&Not available&\\
210&Bowling Green&No seminar series&\\
212**&Southern Mississippi&No seminar series&\\
218&Bucknell&No seminar series&\\
219&Toledo&Economics Department Speaker Series&4\\
221&Kennesaw State&Coles Seminar Series&12 \\
222&Louisiana Tech&No seminar series&\\
223&Rhodes College&No seminar series&\\
224&Central Arkansas&No seminar series&\\
226&Western Kentucky&No seminar series&\\
227&New School&No seminar series&\\
228&Kent State&No seminar series&\\
229&Louisville&No seminar series&\\
230&Texas State&No seminar series&\\
231&Rochester Tech&Gosnell Lecture Series&1\\
232& Central Michigan&No seminar series&\\
233&Northern Iowa&No seminar series&\\
234&Bates College&Seminars&13\\
236&San Jose State&Economics Workshop&4\\
237&U.S. Military Academy&No seminar series&\\
239&Seton Hall&No seminar series&\\
240&Vassar&No seminar series&\\
				\bottomrule
				\end{longtable}
			Not available: The departments' web does not provide information about research seminars in 2018; it provides partial information; or, it is not clear whether seminars are related to the economics department. *= No information on faculty. **= Less than five professors in the department.

\end{document}